# Arabic documents classification using fuzzy R.B.F classifier with sliding window

T.Zaki, M.Amrouch, D.Mammass, A.Ennaji

**Abstract**—In this paper, we propose a system for contextual and semantic Arabic documents classification by improving the standard fuzzy model. Indeed, promoting neighborhood semantic terms that seems absent in this model by using a radial basis modeling. In order to identify the relevant documents to the query. This approach calculates the similarity between related terms by determining the relevance of each relative to documents (NEAR operator), based on a kernel function. The use of sliding window improves the process of classification. The results obtained on a arabic dataset of press show very good performance compared with the literature.

——————— ◆ ———————

## 1 INTRODUCTION

TODAY, Arabic is among the most widely used languages in the world, however, there are few studies on Arabic text classification. In addition, the automatic processing of Arabic seems difficult due to its morphological and syntactic properties [1], [2].

The information retrieval in Arabic is a very interesting domain which poses many problems [3], [4]. The Linguistic approach appears effective, but it is a very laborious task to complete it. This requires specific language tools such as morpho-syntactic and semantic analyzers and thesaurus of the treated language. However, it is important to pass on a semantic level to avoid syntax problems and comparing term by term, hence the need to find methods that assign the correct meaning to words with respect to their context [5], [6]. A state of the art on local semantic similarity measures and the global algorithms for lexical disambiguation based knowledge is detailed in [7]. Several approaches have been proposed, including the standard fuzzy model to the evaluation of the similarity between documents, but most of them ignore the notion of semantic neighborhood. In this respect, we propose an improved version of the standard model for contextual and semantic classification of Arabic documents. To do this, we calculate the similarity between words and local relevance measures using a kernel functions "radial basis".

The use of radial basis modeling is a good solution which consists of using a sliding window to improve the process of classification after obtaining local descriptors according to their semantic relevance.

————————————
- *Taher ZAKI is with the IRFSIC Laboratory, Ibn Zohr University Agadir Morocco and LITIS Laboratory, University of Rouen France.*
- *Mustapha Amrouch is with the IRFSIC Laboratory, Ibn Zohr University Agadir Morocco*
- *Driss Mammass is with the IRFSIC Laboratory, Ibn Zohr University Agadir Morocco*
- *Abdellatif Ennaji is with the LITIS Laboratory, University of Rouen France.*

## 2 PROXIMITY FONCTION

### 2.1 Binary Proximity

the Boolean systems that implement the NEAR operator [8] implicitly use the notion of proximity in their process. The NEAR operator behaves like the AND operator with an additional constraint on the positions of occurrences of the terms concerned specifying a maximum distance between two terms A and B of the query q. For example, if we regard in q, A NEAR 7 B, a system implementing the NEAR operator evaluates this request to the value true if and only if at least one occurrence of the term A is less than 7 words (distance of 7 steps) of at least one occurrence of the word B.

### 2.2 Fuzzy Proximity

Unlike the classical models (logical and vector) that proceed with a global approach (based on the criteria of membership and frequency) of the influence of the occurrences of the terms on the relevance of the document to a query. Therefore, the distribution of the query terms is not involved in the calculation of the score of the relevance of the document to the query.

In models based on fuzzy logic, each term $t \in T$ is associated to an influence function defined on $\mathbb{R}$, bounded support, taking values in $[0, 1]$, symmetric, increasing on $\mathbb{R}^-$ and decreasing on $\mathbb{R}^+$ denoted $\mu_t$ reflecting the degree of belonging document corresponding to the fuzzy set of the term t:

$$\mu_t : D \to [0, 1]$$
$$d \mapsto \mu_t(d) \tag{1}$$

Several types of functions (Gaussian, rectangular functions, features Hanning, triangular, etc ...) can be used. The fuzzy approach makes the notion of proximity fuzzy by assigning a fuzzy interpretation of the NEAR operator.

Indeed, each document is modelled as a finite sequence whose length is equal to l of the text terms T($t_0$, $t_1$,… , $t_{l-1}$) i.e., a function $d : N \to T$ whose definition domain is an interval of N starting at 0. $d^{-1}(t)$ refers to the position set taken by t in document d.

For an occurrence of a term t at position i, the translation G, of the influence function f is used to model the fuzzy proximity where k is a constant determining the width of the area of influence. For example, for a triangular function, the value at the point x is equal to 1 then decreases k at nearby positions until 0. In this case, we can represent the function of influence as follows:

$$f(x) = Max\left(\frac{k-|x|}{k}, 0\right) \quad (2)$$

For example, If we look for A near B, we give a proximity local value to the query NEAR (A, B) in document d by:

$$\mu_{NEAR(A,B)}(d) = Max_{\substack{i \in d^{-1}(A) \\ j \in d^{-1}(B)}} f(x-i) \quad (3)$$

Hence,

$$\mu_{NEAR(A,B)}(d) = Max_{\substack{i \in d^{-1}(A) \\ j \in d^{-1}(B)}} \left( Max\left(\frac{k-|x-i|}{k}, 0\right)\right) \quad (4)$$

The parameter k is integer according to the evaluation context. For example, a value of k = 5 evaluates the proximity in the expression case while k = 100 translated proximity in a paragraph context and so on.

The value we attribute to $\mu_t$ is related to the distance between the two closest occurrences of the two terms A and B in the document. The maximum value is reached when the value of |j-i| is minimum, i.e. equal to 1 because A and B may not appear in the same position. Consequently, we necessarily have j≠i. Therefore, the minimum value is reached when there is an instance of A that is near a B instance in the document. For more details see [9], [10].

### 2.3 Local Relevance Of a Term Relative to a Document

To compare a term and a document, the function $\mu_t^d$ calculates the degree of relevance for each term t of the query q in all possible positions x in d.

The positions x are defined by positive integers as well as by negative ones since the influence of terms extends either side of the their occurrence positions which overflow either before the start of the document or after it has ended.

$$\mu_t^d(x) = Max_{i \in d^{-1}(t)}\left( Max\left(\frac{k-|x-i|}{k}, 0\right)\right) \quad (5)$$

### 2.4 Relevance Of a Query in Relation to a Document

The classical Boolean model considers the request as a series of conjunctions and disjunctions of terms. Thus the assumption that the co-occurrence model there rested. Therefore the local relevance of the q query follows the same logical diagram between the respective relevance of the query terms. The logical operators are the classical operators (AND, OR). The local relevance of q to:

for example, q=A and B,

$$\mu_{A\ OR\ B}^d(x) = max\left(\mu_A^d(x), \mu_B^d(x)\right) \quad (7)$$

and, q=A or B,

$$\mu_{A\ AND\ B}^d(x) = min\left(\mu_A^d(x), \mu_B^d(x)\right) \quad (6)$$

The Relevance to the document is generalized in a natural way by an aggregation of the results obtained in all possible positions.

$$score(r, d) = \sum_{x \in [0, N-1]} \mu_r^d(x) \quad (8)$$

Thus, the similarity is obtained by the normalization of all scores by the cardinality of the fuzzy set $d^{-1}$.

$$Sim(r, d) = \frac{\sum_{x \in [0, N-1]} \mu_r^d(x)}{N} \quad (9)$$

The choice of terms is made simply from a correspondence according to the form of the keywords (lemmas or stems) of the document.

## 3 SYSTEM ARCHITECTURE

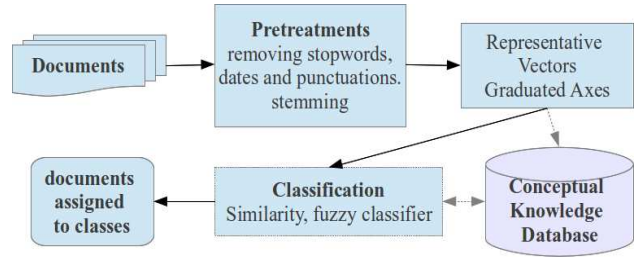

Fig. 1. fuzzy radial basis system with sliding window

The pre-processing phase aims at firstly applying a noise filtering (elimination of empty words, punctuation, date...) to the whole text followed by a morphological analysis (lemmatization, stemming) and then a filtering of extracted terms. This treatment is necessary due to changes in the way in which the text can be represented in Arabic. The preparation of the text consists of the following steps:
1. The text files are converted to UTF-16 encoding.
2. Punctuation marks, diacritics and stop-words are removed.
3. Arabic text normalization consist of transforming some characters in standard form as "إ , أ , آ" to "ا" and "يء , ىء" to "ئ" and "وء" to "و".
4. the terms stemming is performed using the Khoja stemmer [11].

Thereafter, we will proceed to the phase of document representation. This phase eliminates insignificant and irrelevant terms of the considered fields. Then we distinguish between the "descriptor" terms and "equivalent terms." At the end of this phase, we obtain a graduated vector (axis) whose points correspond to the positions of descriptors and their equivalent terms which will be used by the fuzzy classifier in order to assign the appropriate category.

### 3.1 Terms Weighting

Unlike the classic model based on a vector representation whose elements are the frequencies of appearances in documents, or any other statistics measures that refer to this modelling. The fuzzy model calculates the extent to which a term or a query belongs to a document. The result is a vector whose characteristics are the local semantic relevance of the terms. We improved the model [9] using a radial basis modeling to take into account the semantics of the adjacent terms that seems absent in this model. Indeed, a term which has a semantically rich proximity in a document is often relevant to characterize its content.

Starting from this idea, we have proposed a new measure of relevance based on the classic model that takes into consideration the notion of close proximity of terms.

### 3.2 The Radial Basis Fuzzy Proximity

The fuzzy proximity with a radial basis function relies on the determination of supports in the representation space E. However, in contrast to the classical fuzzy model, it may correspond to fictional forms which are a combination of the traditional fuzzy proximity values which we call prototypes. They are associated with a zone of influence defined by a distance (Euclidean, Mahalanobis...) and a radial basis function (Gaussian, exponential,...). The discriminating function g of RBF fuzzy proximity with one output is defined by the distance between the input form of each prototype and the linear combination of the corresponding radial basis functions:

$$g(x) = w_0 + \sum_{i=1}^{N} w_i \, \phi(d(x, \sup_i)) \quad (10)$$

While $d(x, \sup_i)$ is the distance between the input x and the support $\sup_i$, $\{w_0, ..., w_N\}$ are the combination weights and $\phi$ the radial basis function.

Learning this type of model can be done in one or two stages. In the first stage, a gradient method is used to adjust all the parameters by minimizing an objective function based on a criterion such as least-squares. In the second stage, a first step is to determine the parameters related to the radial basic functions (position of the prototypes and zones of influence). To determine the centres, unsupervised classification methods are often used. In a second step, the output layer weights can be learned by different methods such as pseudo-inverse or gradient descent.

In the case of a two-stage learning, the RBF fuzzy proximity functions have so many advantages. For example, the separated learning of the radial basic functions and their combination makes learning fast, simple and avoids the problems of local minima (local and global relevance), the prototypes of the Fuzzy proximity function - RBF represent the distribution of examples in the representation space E (terms). In addition multi-class problem management is simpler with fuzzy proximity functions RBF. The modeling of RBF fuzzy proximity is both discriminating and intrinsic. Indeed the layer of radial basic functions corresponds to an intrinsic description of the training data, then the output combination layer seeks to discriminate different classes.

## 4 OUR RBF FUZZY PROXIMITY MODEL WITH SLIDING WINDOW

In this model the radial basic function associated to the zone of semantic influence is adjusted according to the sliding window size, so as to represent all terms close to the term t according to the used measure of conceptual matching (pairing). This new approach brings semantic knowledge to the already existing model. In this case the local relevance [9] of a term t at position x in a document becomes:

$$\mu_t^d(x) = Max_{i \in d^{-1}(t)} \left( Max \left( \frac{k_f - |x - i|}{k_f}, 0 \right) \right) \quad (11)$$

Where $k_f$ is the sliding window size. The advantage, this model does not need a basic vocabulary to identify terms in order to assign them a weight, since the identification of terms is made simply from a query-document matching according keywords form (lemmas or radicals) of the document. Figure 2 below shows the procedure for calculating the new relevance:

The Discriminating radial basic function of the rele-vance with one output is defined by the distance between the input form to each of the prototype and the linear combination of the corresponding radial basic functions:

$$\mu_t^d(x) = \mu_{0t}^d(x) + \sum_{i}^{k_f} \mu_t^d(i) * \varphi\left(\mu_{t_i}^d\right) \quad (12)$$

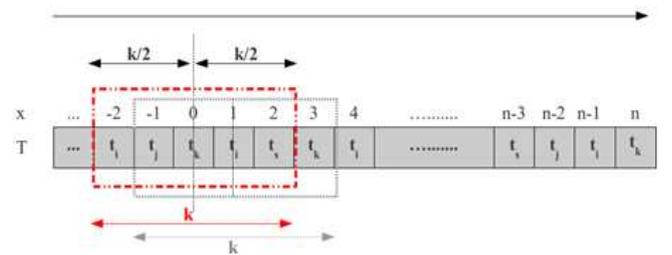

Fig. 2. The radial basis fuzzy process with sliding window.

This model corresponds to an intrinsic description ad-jacent data. The output form seeks to discriminate different labels. The chosen radial basic function is a Gaussian:

$$\varphi(x) = \frac{1}{\sigma\sqrt{2\pi}} e^{\frac{-(x-\mu)^2}{2\sigma^2}} \quad (12)$$

$\mu_{Ot}^d(x)$ the initial relevance value of the term $t$ to the document

$\mu_{t_i}^d(i)$ the relevance value of the term $t_i$ neighbour to t at the position $i$ to the document $d$

$\mu$ the relevance average of the neighbour terms to $t$ according to the sliding window

$\sigma$ the relevance variance of the neighbour terms to $t$ according to the sliding window

A similarity threshold is necessary to characterize the set of terms semantically close to t, it amounts to determine empirically an optimal value of similarity from which the term ti is considered semantically close to t. We choose to use the standard deviation to measure proximity and the dispersion of the data set in the vicinity of the term t. More it is low, most values are grouped around the average, more the neighbourhood data are homogeneous.

## 5 RESULTS

To validate this new approach, we tested it on a varied corpus of 5000 documents electronic press extracted from sites (http://www.aljazeera.net, http://www.alarabiya.net).

Tables 1 and 2 show different results for each measure. These results are expressed through the recall and precision criteria. In particular, they show the relevance of using radial basic functions which greatly improves the measures performance with which they are combined.

TABLE 1
STANDARD FUZZY PROXIMITY RESULTS

| Corpus | recall | precision |
|---|---|---|
| Economy | 0,61 | 0,63 |
| Politic | 0,60 | 0,68 |
| Sport | 0,70 | 0,74 |

TABLE 2
RBF STANDARD FUZZY PROXIMITY WITH SLIDING WINDOW RESULTS

| Corpus | recall | precision |
|---|---|---|
| Economy | 0,69 | 0,75 |
| Politic | 0,67 | 0,77 |
| Sport | 0,78 | 0,84 |

## 6 DISCUSSION AND CONCLUSION

The semantic proximity between words must be highlighted when we deal with complex documents such as texts in Arabic. For this purpose, it is essential to broaden our reflection to the adapted representation models to the nature of our resources. For this, we studied the research model based on the proximity of terms based on the classic fuzzy model. This approach is based on the assumption that most terms occurrences of a query are close in a database document, more this document is relevant to this query, This can partially solve the problems caused by the complex or compound words which may also be an interesting track, since long concepts are often less ambiguous.

However, this model does not consider the notion of terms semantics, since it is limited by the presence of co-occurrence relations of the terms, also does not take into account the semantic links which may exist between the query terms and those of the document.

The integration of a semantic measure between terms in this model is needed. For this reason, we have introduced our radial basis contribution to formalize the adaptation of the model based on the semantic fuzzy proximity concept to the needs of the semantic pairing.

The advantage of this model is that it does not need a preliminary glossary to identify terms in order to assign them a weight, since the identification of terms is made simply from a query-document matching according the shape of document key words (lemmas or radicals).

The integration of the semantic vicinity concept and radial basis functions improves significantly the performance of the classical measures, especially for the Arabic language, which remains our ultimate goal.

**Taher Zaki** received the DESA degree in Computer Science from Ibn Zohr University, and now is a PhD student at the same University, Faculty of Sciences, in the " images pattern recognition systems intelligent and communicating " Laboratory , under the supervision of Prof. Driss Mammass. His research interests systems of information retrieval, text indexing and archive of documents.

**Mustapha Amrouch** received the PhD degree in computer science from Faculty of Sciences, University Ibn Zohr, Agadir, Morocco, in 2012, on automatic recognition of printed and handwritten Amazigh characters, texts and documents. He is currently a researcher of the IRF-SIC Laboratory. His current research interests include pattern recognition, image analysis, indexing and archiving of documents and automatic processing of natural languages.

**Driss Mammass** is professor of Higher Education at the Faculty of Sciences, University Ibn Zohr, Agadir Morocco. He received a Doctorat in Mathematics in 1988 from Paul Sabatier University (Toulouse - France) and a doctorat d'Etat-es-Sciences degrees in Mathematics and Image Processing from Faculty of Sciences, University Ibn Zohr Agadir Morocco, in 1999. He supervises several Ph.D theses in the various research themes of mathematics and computer science such as remote sensing and GIS, digital image processing and pattern recognition, the geographic databases, knowledge management, semantic web, etc. He is currently Director of High School of Technology Agadir and the head of IRF-SIC Laboratory (Image Reconnaissances des Formes, Systèmes Intelligents et Communicants) and an unit of formation and research in doctorat on mathematics and informatics.

**Abdellatif Ennaji** has been an associate professor at the University of Rouen since 1993. He received his Ph.D. from the University of Rouen in 1993 in the fields of machine learning and pattern recognition. His major scientific interest include incremental technics for statistical and hybrid machine learning, data analysis and clustering. The main applications of these activities concern pattern recognition problems and Arabic text mining and recognition. Dr Ennaji has coauthored over 80 publications.